\pdfoutput=1
\documentclass[11pt,a4paper]{article}
\usepackage[hyperref]{paper_arxiv}
\usepackage{times}
\usepackage{latexsym}

\usepackage{microtype}
\usepackage{inconsolata}

\usepackage{multirow}
\usepackage{graphicx}
\usepackage{url}
\usepackage{amsmath}
\usepackage{amssymb}
\usepackage{cprotect}
\usepackage{enumitem}

\aclfinalcopy % Uncomment this line for the final submission
\title{Layout Generation Agents with Large Language Models}

\author{%
	\textbf{Yuichi Sasazawa and Yasuhiro Sogawa} \\
	Hitachi, Ltd. Research and Development Group \\ 
	\texttt{\{yuichi.sasazawa.bj, yasuhiro.sogawa.tp\}@hitachi.com}
}

\begin{document}
	
%\maketitle
\maketitle\thispagestyle{plain}
\begin{abstract}
	
	In recent years, there has been an increasing demand for customizable 3D virtual spaces. Due to the significant human effort required to create these virtual spaces, there is a need for efficiency in virtual space creation. While existing studies have proposed methods for automatically generating layouts such as floor plans and furniture arrangements, these methods only generate text indicating the layout structure based on user instructions, without utilizing the information obtained during the generation process. In this study, we propose an agent-driven layout generation system using the GPT-4V multimodal large language model and validate its effectiveness. Specifically, the language model manipulates agents to sequentially place objects in the virtual space, thus generating layouts that reflect user instructions. Experimental results confirm that our proposed method can generate virtual spaces reflecting user instructions with a high success rate. Additionally, we successfully identified elements contributing to the improvement in behavior generation performance through ablation study~\footnote{Code is available at \url{https://github.com/ckdjrkffz/layout-agent}}.
		
\end{abstract}

	\begin{figure*}[t]
	\centering
	\includegraphics[clip, width=15.5cm]{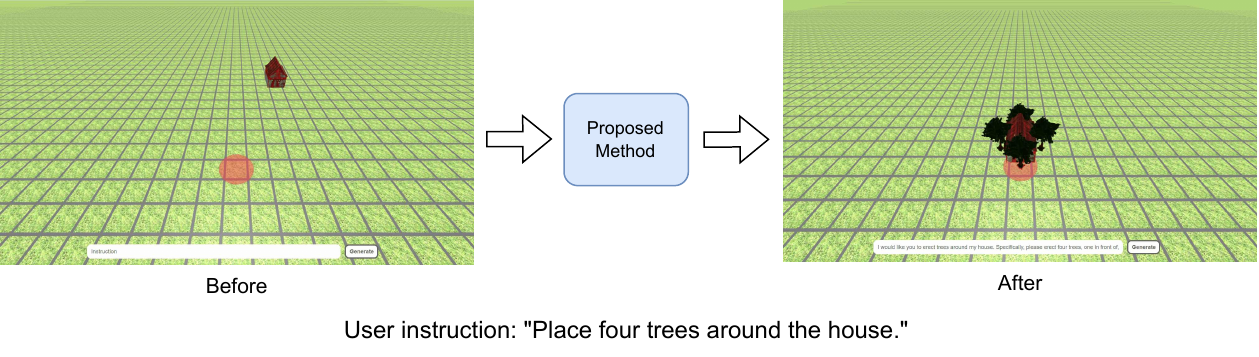}
	\caption{The overview of the task addressed in this study. When the user inputs the instruction text, the proposed method automatically places the object in the appropriate position in sequence, thereby generating a layout that reflects the user's instruction. The gray lines in the image are a grid that conveys coordinate information to the model.}
	\label{fig:task_overview}
\end{figure*}

\section{Introduction}

%Background
In recent years, demand for customizable 3D virtual spaces has been increasing for a wide variety of applications, such as designing room and town layouts, developing electronic games and metaverse, and training AI agents. However, creating virtual spaces for these applications requires manual design, scanning of real environments, and hard-coded rules, all of which require significant human effort~\cite{yang2023holodeck}. Therefore, there is a need for more efficient methods of creating virtual spaces.

%Existing Research
Existing studies have proposed methods for automatic generation of layouts such as house floor plans and indoor furniture arrangements~\cite{Paschalidou2021NEURIPS, feng2023layoutgpt, wen2023anyhome, yang2023holodeck}. However, these methods only generate text indicating the layout structure in response to input user instructions, and do not take advantage of the information in the generation process. In particular, they lack the perspective of utilizing the visual information of the virtual space obtained during the generation process. In addition, most of the existing studies are specialized for generating house floor plans and furniture placement, and do not take into account the application to general purposes. Furthermore, while these methods can create new virtual spaces, they are not suitable for receiving instructions and modifying existing spaces.

%Proposed Method
Against the background of these issues, we propose an agent-operated layout generation system using a large language model (LLM) and verify its effectiveness. In other words, the LLM manipulates agents to place objects one by one in the virtual space to create a layout that reflects user instructions. This approach is inspired by existing research that uses LLMs to manipulate agents in open-ended environments such as Minecraft~\cite{wang2023voyager, lifshitz2023steve1, wang2023jarvis1}. We believe that such an interactive method that takes into account the state of the virtual space enables generic, domain-independent layout generation. Specifically, we use the multimodal language model GPT-4V~\cite{openai2023gpt4} as our action generation model. In each step, the action generation model is given the following information: the agent's action history, and the current state of the virtual space (in image and text format). As output, the model generate the next action to be taken by the agent in JSON format text. The agent changes the state of the virtual space by executing the action in the virtual space. This process is repeated, and terminated when the action generation model determines that the user instructions have been achieved. Objects placed in the virtual space are generated using a 3D object generation model, Shap-E~\cite{jun2023shape}. This allows the appropriate objects to be generated when the action generation model indicates object placement, without the need to prepare a large number of 3D objects in advance.

%Experimental Results
We conducted experiments in a virtual environment to validate the effectiveness of our proposed method. As a result, we confirmed that the proposed method can generate layouts that reflect user instructions with a high success rate. In addition, we conducted ablation studies to clarify which elements of the proposed method contribute to the performance improvement in action generation. As a result, we found that it is difficult for the LLM to grasp the space state only by images, and the input of bounding box text indicating the position of each object instead contributes significantly to the performance improvement of action generation. Performance was also improved by input of past action history, explanation of action intention by using chain-of-thought (CoT)~\cite{wei2023chainofthought}, and specification of move location by absolute position.

\section{Method}
\label{sec:method}

\subsection{Overview}

Figure~\ref{fig:task_overview} shows the task addressed in this study. When a user inputs an instruction text (e.g., ``Place four trees around the house.''), the GPT-4V driven agent performs the operation on behalf of the human and creates a layout that reflects the instruction text.

\begin{figure*}[t]
	\centering
	\includegraphics[clip, width=15.8cm]{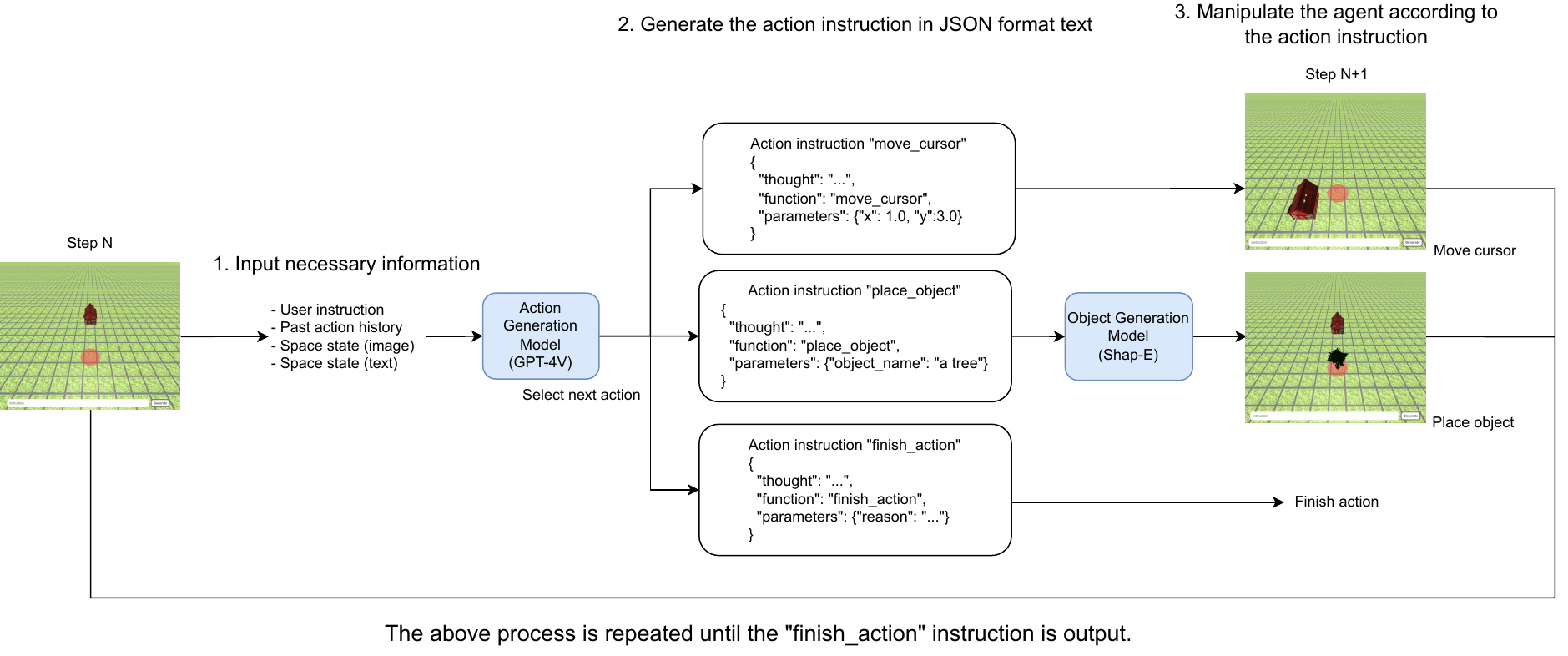}
	\caption{The overview of the algorithm of the proposed method. At each step, necessary information is given to the action generation model (GPT-4V), and the model generates JSON text indicating the next action to be taken by the agent. This process is repeated until the ``finish\_action'' instruction is output, i.e., until the action generation model determines that it has generated a layout that satisfies the user's instruction. Objects to be placed in the virtual space are automatically generated using the 3D object generation model (Shap-E).}
	\label{fig:method_overview}
\end{figure*}

\begin{figure*}[t]
	\centering
	\includegraphics[clip, width=15.8cm]{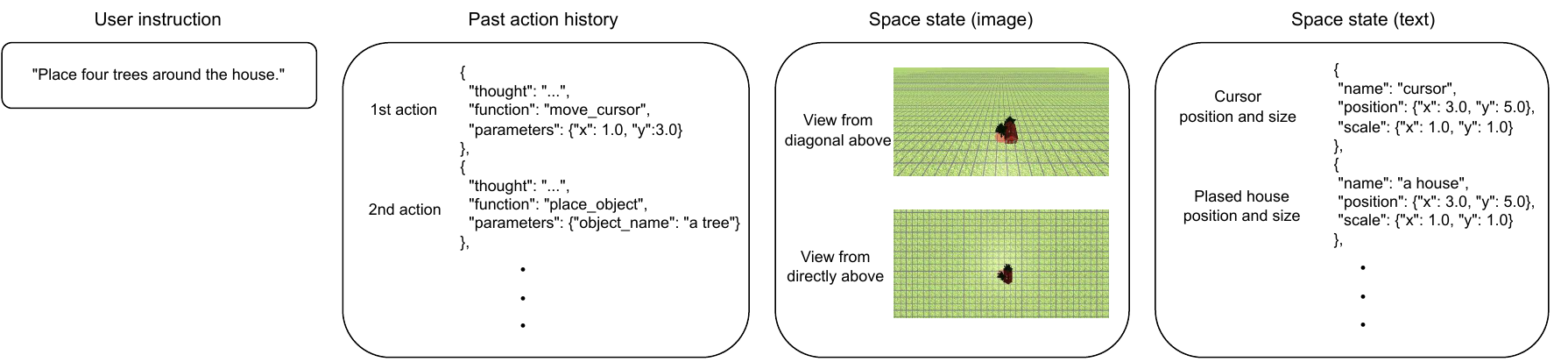}
	\caption{Information to be given to the action generation model}
	\label{fig:model_input}
\end{figure*}

Figure~\ref{fig:method_overview} shows a overview of the algorithm of the proposed method. There is a red circular cursor indicating the current selection position in the virtual space, and the agent manipulates this cursor to generate the layout. The action generation model receives the necessary information shown in Figure~\ref{fig:model_input} at each step, and specifies the next action by generating JSON format text containing the function name and the arguments of the function. The functions that can be used by the action generation model in this study are: 1. \textbf{move\_cursor function} which moves the cursor to the desired position, 2. \textbf{place\_object function} which places the specified object at the current position of the cursor, 3. \textbf{finish\_action function} which determines that user instructions have been fulfilled and terminate execution. If the move\_cursor function or the place\_object function are selected, the agent changes the state of the virtual space by performing the selected action. If the finish\_action function is selected, execution is terminated at that point.

\subsection{Output of the action generation model}

\paragraph{Generation format}: The action generation model indicates the next action by generating JSON format text with the following parameters for each step.

\begin{itemize}[noitemsep,topsep=0.8pt]
	\setlength{\parskip}{1.0pt}
	\setlength{\itemsep}{1.0pt}
	
	\item \textbf{thought}: Natural language text describing the current situation and the next action to be taken by the agent. This text is not used directly in the agent's action, but it is expected to improve performance by generating this text first, which causes a CoT effect. It can also be used by the user to check the reason for unnatural action instructions later when they are generated.
	
	\item \textbf{function}: The name of the function to use. This name must be one of ``move\_cursor'', ``place\_object'', or ``finish\_action''.
	
	\item \textbf{parameters}: Arguments corresponding to the parameters of each function. Details are described below.
	
\end{itemize}

\paragraph{Function type}: Details of the three types of functions used by the action generation model are given below.

\begin{itemize}[noitemsep,topsep=0.8pt]
	\setlength{\parskip}{1.0pt}
	\setlength{\itemsep}{1.0pt}
	
	\item \textbf{move\_cursor}: Moves the cursor to the specified position. The parameters are the 2D coordinates (``x'', ``y'') of the destination.
	
	\item \textbf{place\_object}: Places the specified object at the current cursor position. The parameter is a text that describes what kind of object should be created (``object\_name''). The object to be placed is created by supplying the ``object\_name'' parameter text to the object generation model. Once an object has been generated, it is saved, and when the same object placement is indicated again, the previously generated object is reused.
	
	\item \textbf{finish\_action}: The execution terminates when it judges that the user instruction has been completed. The parameter is a text that explains to the user why the execution was terminated (``reason'').
	
\end{itemize}

\subsection{Input to the action generation model}

For each step, the following information is given to the action generation model as input.

\begin{itemize}[noitemsep,topsep=0.8pt]
	\setlength{\parskip}{1.0pt}
	\setlength{\itemsep}{1.0pt}
	
	\item \textbf{Common Prompt}: Text describing the task, available function choices, output format, etc. This text is always common.
	\item \textbf{User Instruction}: Text that instructs what kind of layout should be created. This text is common to each step.
	\item \textbf{Action History}: Text indicating the agent's previous action history. In other words, it is a list of JSON format texts (including thought texts) indicating agent actions generated by the action generation model in past steps, arranged in the order of execution.
	\item \textbf{Space State (Image)}: Information on an image showing the current space state of the virtual space. Two views, one diagonally above and one directly above the cursor, are use to help the model grasp the space conditions. A grid consisting of gray lines aligned at regular intervals is displayed on the space to help the model understand the appropriate distance to move when using the move\_cursor function.
	\item \textbf{Space State (Text)}: A Bounding box format text indicating the current position (2D coordinates) and size for each object in virtual space (cursor, pre-installed objects, and objects placed by the action generation model).

\end{itemize}

\section{Experiment}

\subsection{Purpose of the Experiment}

The purpose of the experiment is to evaluate whether the proposed method can generate appropriate layouts for user instructions. In addition, an ablation study is performed to verify which elements of the proposed method contribute to improving the performance of action generation. The number of samples used in this experiment is very small and therefore insufficient for a quantitative evaluation. However, the difference in scores among the methods was large, and the superiority of the methods can be judged relatively clearly, so we believe that certain insights were gained.

\subsection{Experimental setup}

GPT-4V (``gpt-4-vision-preview'')~\cite{openai2023gpt4} of OpenAI API was used for the behavior generation model, and Shap-E~\cite{jun2023shape} was used for the object generation model. The temperature of the behavior-generating model was set to $0.1$. The resolution of the input image was set to $512 \times 256$. A-Frame~\footnote{\url{https://aframe.io/}} was used to construct the virtual space environment, and the initial state of the virtual space for the experiments was manually created.

The evaluation was conducted using the five user instruction texts shown in Table~\ref{tab:prompt}. We conducted the experiment three times for each instruction, and manually evaluated each of the obtained final space states on a scale of 1 to 3, and report the average value. The evaluation was made by looking at the state of the virtual space in terms of whether an appropriate layout was generated in response to user instructions. However, the naturalness of the order of the agents' actions and factors that the action generation model cannot control, such as the orientation and size of objects, were not considered in the evaluation.

In the experiment, we compared the main proposed method (Method 1) and the ablation study method (Methods 2 to 6), which did not use some of its components, for a total of six methods. Details of each method are given below.

\begin{itemize}[noitemsep,topsep=0.8pt]
	\setlength{\parskip}{1.0pt}
	\setlength{\itemsep}{1.0pt}
	
	\item \textbf{Method 1 (all)} : This is the complete proposed method, which includes all the elements described in Section~\ref{sec:method}.

	\item \textbf{Method 2 (w/o history)} : The agent's past action history is not input into the action generation model.
	
	\item \textbf{Method 3 (w/o status image)} : The space state (image) is not input into the action generation model.
	
	\item \textbf{Method 4 (w/o status text)} : The space state (text) is not input into the action generation model. In this setting, we can verify the ability to move objects to the coordinates where they should be placed using only the grid that appears in the image. However, only the information on the current position and size of the cursor at each step is input, as it is necessary to specify the coordinates of the destination.
	
	\item \textbf{Method 5 (w/o CoT)} : CoT is not used. That is, when the action generation model generates the JSON format instruction text, it does not generate ``thought'' parameter.
	
	\item \textbf{Method 6 (w/o absolute pos)} : When using the move cursor function (move\_cursor), the destination coordinates is specified not by absolute position, but by relative position that indicates how far the cursor should move in x and y direction from the current position. For example, when moving from $(x,y)=(3,5)$ to $(x,y)=(4,7)$, $(x,y)=(4,7)$ is specified when using absolute position, but $(x,y)=(1,2)$ is specified instead when using relative position.

\end{itemize}

\begin{table}[t]
	\centering
	\caption{Five instruction texts used in the evaluation}
	\scalebox{0.85}{
		{\tabcolsep=3.0pt
			\begin{tabular}{cc} \hline
				No & User instructions \\ \hline
				1 & \parbox{8.0cm}{Place one tree in the immediate vicinity of the house.} \\ \hline
				2 & \parbox{8.0cm}{Place three trees on the left side of the house.} \\ \hline
				3 & \parbox{8.0cm}{Place trees around the house. Specifically, place four trees, one in front of, to the left of, to the right of, and behind the house.} \\ \hline
				4 & \parbox{8.0cm}{Place one bed in the room. Place it near the wall on the left side.} \\ \hline
				5 & \parbox{8.0cm}{Place one bed and one bookshelf against the wall of the room and one small desk in the center of the room.} \\ \hline
			\end{tabular}
		}
	}
	\label{tab:prompt}
\end{table}

\begin{table}[t]
	\centering
	\caption{Experimental results. We report the average of the human rating on a scale of 1 to 3 for each of the proposed method (Method 1) and the ablation studies without some of the components of the proposed method (Methods 2 to 6).}
	\scalebox{1.0}{
		{\tabcolsep=2.0pt
			\begin{tabular}{lc} \hline
				\multirow{2}{*}{Method} & Evaluation Score \\
				& (1.0 - 3.0) \\ \hline
				Method 1 (all) & \underline{2.9} \\
				Method 2 (w/o history) & 2.7 \\
				Method 3 (w/o status image) & \textbf{3.0} \\
				Method 4 (w/o status text) & 1.3 \\
				Method 5 (w/o CoT) & 2.6 \\
				Method 6 (w/o absolute pos) & 1.7 \\ \hline
			\end{tabular}
		}
	}
	\label{tab:main_result}
\end{table}

\begin{figure}[t]
	\centering
	\includegraphics[clip, width=7.8cm]{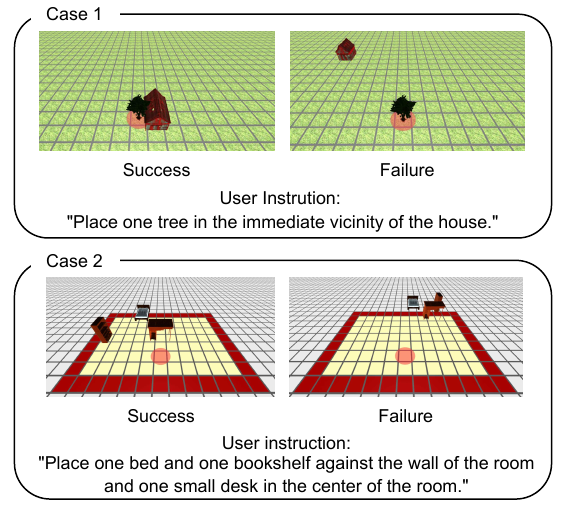}
	\caption{Sample of generated results. Case 1 shows an example of placing a tree near a house on a meadow, and Case 2 shows an example of placing furniture in a room (the brown frame is a wall).}
	\label{fig:example}
\end{figure}

\subsection{Result}

Table~\ref{tab:main_result} shows the experimental results and Figure~\ref{fig:example} shows samples generated by our method. Method 1 (all) achieves an extremely high success rate, confirming that the proposed method works properly. Method 3 (w/o status image) achieves almost the same performance as Method 1 (all). This result indicates that the proposed method does not require representation in image format in order to understand space conditions, and bounding box-style text format is sufficient. However, it is possible that there are elements that are difficult for the model to grasp only with text information, such as whether the orientation and size of the placed objects are appropriate in relation to the surrounding environment, and the development and evaluation of a method that can manipulate these elements is a future work. Also, it has been reported that displaying a dot matrix and text indicating coordinates on the image enhances the model's ability to recognize space~\cite{lei2024scaffolding}, and the incorporation of this method is also the future work.

Method 2 (w/o history) and 5 (w/o CoT) performed slightly worse than Method 1 (all). In particular, Method 2 (w/o history) often performed inconsistent actions because it was not given the previous action history. For example, when ``placing a tree near the house'' was required, it frequently moved several times alternately between two specific locations without being able to decide where to place the tree. Method 4 (w/o status text) had the lowest performance and did not work properly in most cases. This indicates that it is difficult for the action generation model to properly specify the coordinates of the destination using only the grid on the image. Method 6 (w/o absolute pos) also performed significantly worse, suggesting that absolute position is a more suitable means for the LLM to specify the location of the destination than using relative position.

In short, the contribution of each element in the generation of the agent's action can be summarized as follows:

\begin{itemize}[noitemsep,topsep=0.8pt]
	\setlength{\parskip}{1.0pt}
	\setlength{\itemsep}{1.0pt}
		
	\item Space information (images): Performance is not affected even when not used.

	\item Action history, CoT: If not used, performance is reduced to some extent, but the system can generate layouts that reflect user instructions with a certain probability.
	
	\item Space information (text), use of absolute position: If not used, performance is significantly reduced, and in most cases the user instructions cannot be achieved.
	
\end{itemize}

\section{Conclusion}

In this study, we proposed a system for generating layouts on virtual spaces using a LLM. Experimental results show that the proposed system can generate virtual spaces that reflect user instructions with a high success rate. Through ablation studies, we succeeded in identifying factors that contribute to improving the performance of action generation.

Future tasks include the following:

\begin{itemize}[noitemsep,topsep=0.8pt]
	\setlength{\parskip}{1.0pt}
	\setlength{\itemsep}{1.0pt}
	
	\item Evaluation on a large dataset: The evaluation experiments in this study were conducted on a limited number of sample cases. It is necessary to validate the effectiveness of the proposed method on a larger and more diverse dataset.
	
	\item Addition of functions: The three functions used to instruct the actions in this study alone may limit the range of possible operations. Adding functions, for example, to move, delete, rotate, and resize placed objects, would allow the agent to respond to a greater variety of user instructions.	
	
	\item Faster action generation: The proposed method has a high time cost to generate the final layout because the actions of the agent are generated sequentially at each step. To solve this problem, for example, multiple actions can be performed in a single step.
	
\end{itemize}

\bibliography{reference}
\bibliographystyle{acl_natbib}

\end{document}